\documentstyle[aps,preprint]{revtex}
\tightenlines
\begin{document}

\draft
\title{DNA Torsional Solitons in Presence of localized Inhomogeneities}
\author{Luis Morales Molina$^{(a),(b)}$\thanks{Corresponding author email: lmolina@ff.oc.uh.cu},Owen Pavel Fernandez$^{(a),(c)}$}

\address{$^{(a)}$ Department of Theoretical Physics, University
of Havana, 10400, Havana, Cuba}
\address{$^{(b)}$ Department of Physics, University
of Camaguey, 74560,Camaguey, Cuba}
\address{$^{(c)}$ Department of Physics, University
of Oriente, 90500,Santiago de Cuba, Cuba}

\date{December 23, 2001}

\maketitle

\begin{abstract}
\ In the present paper we investigate the influence of inhomogeneities in the dynamics and stability of DNA open states, modeled as propagating solitons in the spirit of a Generalized Yakushevish Model. It is a direct consequence of our model that there exists a critical distance between the soliton's center of mass and the inhomogeneity at which the interaction between them can change the stability of the open state.Furtherly from this results was derived a renormalized potential function.  
\end{abstract}


\section{Introduction}
\ The origin and dynamics of open states in DNA have been the subject of intense study in the last three decades\cite{En1,Naka1,Naka2,peyrard,yaku0}. To date some nonlinear models have been proposed\cite{yaku0,En2,Yo1,Yo2,Fedyanin,Zhang,Muto1,Volkov} that addresses this issue, which is somehow related to the
transcription or replication processes in this molecule, in terms of solitons and their interactions. The torsional model proposed by Yakushevich\cite{yaku0,En2,Yo1,Yo2,Fedyanin,Zhang,Muto1,Volkov,yaku1} is one of such models, and in this paper we generalize it to take into account the effects of inhomogeneity, which take place in real DNA molecules\cite{yaku2,Muto2}.This problem has been considered before\cite{yaku0,En2,Yo1,Yo2,Fedyanin,Zhang,Muto1,Volkov,yaku1,yaku2,Muto2,salerno} but in the framework of a simpler version of the model; we are not aware that the effect of the inhomogeneities have been studied previously within the full Yakushevich model.
\\ This paper is organized as follows. In section 2 a description of our generalized model and the fundamental equations are presented. In section 3 an analysis of the model are presented. Furthermore, in order to complete this analysis, some energetics consideration are carried out in section 4.
The last section summarizes our results and in it we make some suggestions for further research on the subject.

\section{Description of the model}

\ To study the DNA torsional dynamics in the spirit of the model proposed by Yakushevich\cite{yaku0,En2,Yo1,Yo2,Fedyanin,Zhang,Muto1,Volkov,yaku1}, we consider the DNA molecule as two long elastic and weakly interacting rods that are bound around each other, producing a double helix.
Each rod represents one of the polynucleotide chains of the DNA molecule. For simplicity, we neglect the helical structure, and instead of the double helix we consider only parallel rods each having the form of the straight line\cite{yaku0,En2,Yo1,Yo2,Fedyanin,Zhang,Muto1,Volkov,yaku1}. Assuming that the twist and the bend motions are independent and only taking into account the first of the above motions, we have the mentioned torsional model.
\\ One of the limitations of the model considered above is that in it the effects of inhomogeneities are neglected. The changes in the strength of hydrogen bonds for different pairs of nitrogen bases, i.e.,the inhomogeneous character of the DNA molecule, can be taken into account incorporating to the equations proposed by Yakushevich a new term that represents the presence of such inhomogeneities. Assuming a local character for these bonds, distributed along the double helix in sites denoted by $ z_{n}, n=1,2,..$
\, the DNA torsional dynamics is now described by the following equations:
\begin{mathletters}
\label{todas1}
\begin{eqnarray}
\ I_{1}\frac{\textstyle\partial^{2}\phi_{1}}{\textstyle\partial t^{2}}-K_{1}a^{2}\frac{\textstyle\partial^{2}\phi_{1}}{\textstyle\partial z^{2}}+ k\frac{\textstyle\Delta l}{\textstyle l}[(2R^{2}+l_{0}R)\sin(\phi_{1})-R^{2}\sin(\phi_{1}+\phi_{2})]    \nonumber
\\+M\sum_{n}\delta(z-z_{n})\frac{\textstyle\Delta l}{\textstyle l}[(2R^{2}+l_{0}R)\sin(\phi_{1})-R^{2}\sin(\phi_{1}+\phi_{2})]=0 \label{1a}
\end{eqnarray}
\begin{eqnarray}
\ I_{2}\frac{\textstyle\partial^{2}\phi_{2}}{\textstyle\partial t^{2}}-K_{2}a^{2}\frac{\textstyle\partial^{2}\phi_{2}}{\textstyle\partial z^{2}}+ k\frac{\textstyle\Delta l}{\textstyle l}[(2R^{2}+l_{0}R)\sin(\phi_{2})-R^{2}\sin(\phi_{1}+\phi_{2})]    \nonumber
\\+M\sum_{n}\delta(z-z_{n})\frac{\textstyle\Delta l}{\textstyle l}[(2R^{2}+l_{0}R)\sin(\phi_{2})-R^{2}\sin(\phi_{1}+\phi_{2})]=0 \label{1b}
\end{eqnarray}
\end{mathletters}
\ where $ \Delta l/l=1-l_{0}[(2R
+l_{0}-R\cos\phi_{1}-R\cos\phi_{2})^{2}+(R\sin\phi_{1}-R\sin\phi_{2})^{2}]^{-\frac
12}$ \ and $\delta(x)$ is the Dirac's Delta function of argument
$x$.
\\ In these equations, $\phi_{i}(i=1,2)$ is the rotational angle, $I_{i}$ is the moment of inertia, $K_{i}$ is the rigidity of the longitudinal springs of the $i-th$ chain, $k$ is the rigidity of the transversal springs connecting both chains(weak bonds), M is the rigidity of the springs modeling the local inhomogeneities that connect the chains in the positions labeled by $z_{n}$, $R$ the radius of the chains, $l_{0}$ the maximum separation between chains and $a$ is the characteristic length of the base pair in the double helix(see Fig.1).
As in reference, we can simplify these equations by assuming $l_{0}=0$, which leads to:
\begin{mathletters}
\label{todas2}
\begin{eqnarray}
\ I_{1}\frac{\partial^{2}\phi_{1}}{\partial\ t^{2}}-K_{1}a^{2}\frac{\partial^{2}\phi_{1}}{\partial\ z^2}
\ + kR^2[2\sin(\phi_{1})-\sin(\phi_{1}+\phi_{2})]\nonumber
\\ +MR^{2}\sum_{n}\delta(z-z_{n})
[2\sin(\phi_{1})-\sin(\phi_{1}+\phi_{2})]=0 \label{2a}
\end{eqnarray}
\begin{eqnarray}
\ I_{2}\frac{\partial^{2}\phi_{2}}{\partial\ t^{2}}-K_{2}a^{2}\frac{\partial^{2}\phi_{2}}{\partial\ z^{2}}
\ + kR^2[2\sin(\phi_{2})-\sin(\phi_{1}+\phi_{2})]\nonumber
\\ +MR^{2}\sum_{n}\delta(z-z_{n})
[2\sin(\phi_{2})-\sin(\phi_{1}+\phi_{2})]=0  \label{2b}
\end{eqnarray}
\end{mathletters}

\section{Analysis of the Model}
\ Now we proceed to analyze the Eqs.(\ref{todas2}) in a general way. As it is of common usage. Let us to introduce the traveling wave variable $\xi=z-vt$, where $v$ is a constant. Then we obtain the following system:

\begin{eqnarray}
\ -W_{i}\frac{\partial^{2}\phi_{i}}{\partial\xi^2}+ kR^{2}[2\sin(\phi_{i})-\sin(\phi_{1}+\phi_{2})]&&\nonumber
\\&&+MR^{2}\sum_{n}\delta(\xi-\xi_{n})[2\sin(\phi_{i})-\sin(\phi_{1}+\phi_{2})]=0 \end{eqnarray}
\        with $i=1,2$
\\ where $W_{i}=K_{i}a^{2}-I_{i}v^{2}$ and $\xi_{n}=z_{n}-vt$.
In the last system, we have considered for simplicity the presence
of one inhomogeneity located at the point $z_{0}$.
\\The system  Eqs.(3) is equivalent to the following dynamical system:

\begin{mathletters}
\label{todas4}
\begin{eqnarray}
\frac{\partial\phi_{1}}{\partial\xi}&=&\Phi \label{4a}
\\
\frac{\partial\Phi}{\partial\xi}&=&
\frac{1}{W_{1}}(kR^{2}+MR^{2}\delta(\xi-\xi_{0}))[2\sin(\phi_{1})-\sin(\phi_{1}+\phi_{2})] \label{4b}
\\
\frac{\partial\phi_{2}}{\partial\xi}&=&\Psi \label{4c}
\\
\frac{\partial\Psi}{\partial\xi}&=& \frac{1}{W_{1}}(kR^{2}+MR^{2}\delta(\xi-\xi_{0}))[2\sin(\phi_{2})-\sin(\phi_{1}+\phi_{2})] \label{4d}
\end{eqnarray}
\end{mathletters}

\ For points $\xi\neq\xi_{0}$,the Eqs.(\ref{todas4}) reduces to the system studied in reference\cite{Jorge}.

\begin{mathletters}
\label{todas5}
\begin{eqnarray}
\frac{\partial\phi_{1}}{\partial\xi}&=&\Phi \label{5a}
\\ \frac{\partial\Phi}{\partial\xi}&=& \frac{1}{W_{1}}kR^{2}[2\sin(\phi_{1})-\sin(\phi_{1}+\phi_{2})] \label{5b}
\\ \frac{\partial\phi_{2}}{\partial\xi}&=&\Psi \label{5c}
\\ \frac{\partial\Psi}{\partial\xi}&=& \frac{1}{W_{1}}kR^{2}[2\sin(\phi_{2})-\sin(\phi_{1}+\phi_{2})] \label{5d}
\end{eqnarray}
\end{mathletters}

\ The last equations corresponds to the potential function:

\begin{equation}
\ V(\phi_{1},\phi_{2})=kR^{2}[2(\cos\phi_{1}+\cos\phi_{2})-\cos(\phi_{1}+\phi_{2})] \label{6}\mbox{,}
\end{equation}
\ which has a maxima at the point $\phi_{1}=2n\pi$, $\phi_{2}=2m\pi$, with $n=0,\pm1,\pm2,...$ and $m=0,\pm1,\pm2,...$. the points $\phi_{1}=\left(2n-1\right)\pi$, $\phi_{2}=\left(2m-1\right)\pi$, correspond to local minima, while the points $\phi_{1}=\left(2n-1\right)\pi$, $\phi_{2}=2m\pi$ and $\phi_{1}=2n\pi$,$\phi_{2}=\left(2m-1\right)\pi$ are saddle points. The trajectories joining these critical points correspond to kink solitons\cite{Jorge}.
\ For  $\xi\sim\xi_{0}$ the system has the same soliton solution, now perturbed by the presence of the local inhomogeneities at this point, as we will see later.
\\ Let us analyze the situation $\phi_{1}=-\phi_{2}$. In references\cite{Jorge,yaku3} it has been shown that for this case there exists exact soliton-like solutions. If the distance between the inhomogeneity and the center of mass of the soliton is much greater than the soliton radius, the interaction between these two entities is very weak, and our model reproduces the solution of the mentioned references. The interaction becomes appreciable at a certain critical distance, that we will estimate later below.

\ Let us consider symmetrical chains, i.e.;$I_{1}=I_{2}=I$ and $K_{1}=K_{2}=K$ and only one inhomogeneity. Then we can write Eqs.(\ref{todas2})as:

\begin{mathletters}
\label{todas7}
\begin{eqnarray}
\ \frac {1}{c^{2}}\frac{\partial^{2}\phi_{1}}{\partial\ t^{2}}-\frac{\partial^{2}\phi_{1}}{\partial\ z^{2}}&&
+\frac{1}{\Lambda^{2}}[2\sin(\phi_{1})-\sin(\phi_{1}+\phi_{2})]\nonumber
\\ &&+\frac{Q}{\Lambda^2}\delta(z-z_{0})[2\sin(\phi_{1})-\sin(\phi_{1}+\phi_{2})]=0\ \label{7a}
\\ \frac{1}{c^{2}}\frac{\partial^{2}\phi_{2}}{\partial\ t^{2}}-\frac{\partial^{2}\phi_{2}}{\partial\ z^{2}}
\ &&+\frac{1}{\Lambda^{2}}[2\sin(\phi_{2})-\sin(\phi_{1}+\phi_{2})] \nonumber
\\&&+\frac{Q}{\Lambda^2}\delta(z-z_{0})[2\sin(\phi_{2})-\sin(\phi_{1}+\phi_{2})]=0 \label{7b}
\end{eqnarray}
\end{mathletters}

\ where $\Lambda^{2}=Ka^{2}/kR^{2}$, $Q=M/k$ and $c^2=Ka^2/I$.
\\ To investigate the stability of non-perturbed soliton solution that in our case is\cite{Jorge,yaku3},

\begin{equation}
\phi_{1}(z)=-\phi_{2}(z)=\phi_{K}(z)=4\arctan\left[\exp\left(\frac{z-z_{m}}{2r}\right)\right] \label{8}
\end{equation}

\ where $z_{m}$ is the point that localizes the soliton radius $r$(half width of the soliton) defined
by

\begin{equation}
\ r=\Lambda/2\sqrt{2}\label{9}
\end{equation}

\ We consider small-amplitude excitations around $\phi_{i}(z)$, i.e., we put:
\begin{equation}
\phi_{i}(z,t)=\phi_{i}(z)+f_{i}(z)\exp(\lambda_{i} t),\  \  i=1,2\label{10}
\end{equation}

\ where $f_{i}(z)$ are small. Linearizing in a standard way \cite{Bishop} Eqs.(\ref{todas7}) around $\phi_{i}(z,t)$ we get the following spectral problems for the exited modes:

\begin{mathletters}
\begin{eqnarray}
\ -\frac {d^{2}f_{1}}{dz^{2}}+\frac{\lambda^2}{c^2}\ f_{1}+&&\frac{1}{\Lambda^{2}}\left[2\cos\left(\phi_{K}\right)f_{1}-f_{1}-f_{2}\right] \nonumber
\\&&+\frac{Q}{\Lambda^{2}}\delta\left(z-z_{0}\right)\left[2\cos\left(\phi_{K}\right)f_{1}-f_{1}-f_{2}\right]=0\label{11a}
\\
-\frac {d^{2}f_{2}}{dz^{2}}+\frac{\lambda^2}{c^2}\ f_{2}+&&\frac{1}{\Lambda^{2}}\left[2\cos\left(\phi_{K}\right)f_{2}-f_{1}-f_{2}\right] \nonumber
\\&&+\frac{Q}{\Lambda^{2}}\delta\left(z-z_{0}\right)\left[2\cos\left(\phi_{K}\right)f_{2}-f_{1}-f_{2}\right]=0\label{11b}
\end{eqnarray}
\end{mathletters}

\ We can write the above system as follows:

\begin{eqnarray}
\ \frac {d^{2}F_{\pm}}{dz^{2}}-N_{\pm}F_{\pm}&&+ \frac{4}{\Lambda^{2}}\frac {F_{\pm}}{\cosh^{2}\left(\sqrt{2}\left(z-z_{m}\right)/\Lambda\right)}\nonumber
\\
&&-\frac{2Q}{\Lambda^{2}}\left[1-\frac{2}{\cosh^{2}\left(\sqrt{2}\left(z_{0}-z_{m}
\right)/\Lambda\right)}\right]\delta\left(z-z_{0}\right)=0\label{12}
\end{eqnarray}

\ where $F_{+}$ and $F_{-}$ are symmetric and antisymmetric solutions, i.e.,$F_{\pm}=(f_{1}\pm\ f_{2})/2$,
$N_{+}$,$N_{-}$ corresponds to $F_{+}$, $F_{-}$ and:

\begin{equation}
N_{+}=\frac{\lambda_{+}^{2}}{c^{2}}, \hspace{1cm}\ N_{-}=\frac{\lambda_{-}^{2}}{c^{2}}+\frac{2}{\Lambda^{2}}  \label{13}
\end{equation}

\ We can see that $\chi=z_{0}-z_{m}$ represents the distance between the inhomogeneity and the soliton's center of mass. The soliton is unstable for positive eigenvalues.
For the spectral problem Eqs.(\ref{12}),(\ref{13})there exist two positive eigenvalues corresponding to the discrete spectrum[see Appendix]. The solution corresponding to the first positive eigenvalue:

\begin{equation}
\lambda_{+}=c\sqrt{\frac{2}{\Lambda^{2}}+\frac{Q^{2}}{\Lambda^{4}}
\left[\frac{2}{\cosh^{2}(\sqrt{2}\chi/\Lambda)}-1\right]^{2}}\label{14}
\end{equation}

\ is

\begin{equation}
\ f_{1}=f_{2}=F_{+}(z)=\left\{\begin{array}{c}
\ A \frac{
\textstyle\exp\left(-\frac{\textstyle Q}{\textstyle \Lambda^{2}}
\left[\frac{\textstyle 2}{\textstyle \cosh^{2}(\sqrt{2}\chi/\Lambda)}-1\right]\textstyle\left(z-z_{0}\right)
\right)}{\textstyle\cosh(\sqrt{2}(z-z_{m}))}\   \hspace{0.5cm}\      z>z_{0}
\\
\\ \ A \frac{
\textstyle\exp\left(\frac{\textstyle Q}{\textstyle\Lambda^{2}}
\left[\frac{\textstyle 2}{\textstyle\cosh^{2}(\sqrt{2}\chi/\Lambda)}-1\right]\textstyle\left(z-z_{0}\right)
\right)}{\textstyle\cosh(\sqrt{2}(z-z_{m}))}\hspace{0.5cm}         z<z_{0}
\end{array}   \label{15}
\right.
\end{equation}
\ where $A$ is determined from the normalization condition.
\\
\ For the other positive eigenvalue:

\begin{equation}
\lambda_{-}=c\frac{Q}{\Lambda^{2}}\left|\left[\frac{2}{\cosh^{2}\left(\sqrt{2}\chi/\Lambda\right)}-1\right]\right|    \label{16}
\end{equation}

\ we have for $f_{1}=-f_{2}=F_{-}(z)$ the same expressions Eq.(\ref{15}).
\ As we can see from the equations above if we put in these $Q=0$ and $z_{m}=0$, the results coincide with those reported in reference \cite{Jorge} for these solitons.
\ From the condition of instability($\lambda>0$), the equation Eq.(\ref{A2}) and the condition  Eq.(\ref{A6}) we find that for only distances smaller than a critical distance the inhomogeneity can affect the structural stability of the open state, i.e., that for distances(between the propagating soliton and the inhomogeneity) greater than such a critical distance, the interaction only influences the dynamics of its motion. This fact can be explained because for  greater distances the inhomogeneity has not influence on the discrete states of the eigenvalue problem for the stability analysis, namely its solution reduces to the translational mode which is stable.
 This change in the stability can only to be due to a releasing of energy, i.e, a rupture of the stronger bond.
\\ The critical distance at which a rupture of the stronger bond occur, can be obtained from the Eq.(\ref{A9}). This can written as a function of soliton's radius as:

\begin{equation}
\ \chi=d_{crit}=1.606 r    \label{17}
\end{equation}

\ The existence of a critical distance starting from which the bond is broken giving place to new implications no cleared up to now. For that reason we make in the following  a simple energetic analysis.

\section{Some energetics considerations}

\ In this section we briefly discuss some aspects concerning to
the energetic contributions that appear in the presence of the
inhomogeneities and it is a proposed field in order to model the
motion of solitons as a point-particles as is usually done
\cite{Kivshar1,Kivshar2,kaberman,sanchez}. The total energy for
this mechanical model without to take into account the
contribution due to inhomogeneities is given by the following
expression:

\begin{equation}
\ E_{0}=E_{K}+E_{g}+E_{T}   \label{18}
\end{equation}

\ where $E_{g}$, $E_{T}$ are the energies due to elastic force between the cylinders of different chains and the torsional force between the cylinders of the same chain respectively. $E_{K}$ is the corresponding kinetic energy.
\\ In the presence of inhomogeneities $E$ has the form
\begin{equation}
\ E=E_{0}+V        \label{19}
\end{equation}

\ where the term

\begin{equation}
\ V=\frac{MR^{2}\Delta \phi^{2}}{2}    \label{20}
\end{equation}
\ takes into account the difference between the strength of weak
and strong bonds.
\\ From the energetic point of view the traveling solitary wave must give an additional energy for breaking such a bond. Neglecting the dissipative effects and making the assumption that the rupture is produced when the solitary wave center is separated from the inhomogeneity by the critical distance then the rupture is possible if the effective kinetic energy has a value greater than the potential energy at the critical distance.
\\ If this hypothesis is true it is possible to model the motion of solitons as free particles in a potential field $V$ with an effective kinetic energy $E_{0}$. This function $V$ is a renormalized potential, which picking up the main effects due to the interaction with the inhomogeneity.
\\ In order to build the potential function we have taken into account that for distances lower than the critical distance between the center of soliton and the inhomogeneity the elastic energy becomes zero because of the rupture of the bond at this distance. In a general form the potential function is:

\begin{equation}
\ V(\xi)=\left\{\begin{array}{c}
\frac{\textstyle\ MR^2}{\textstyle\ 2}[(\phi_{2u}(\infty)-\phi_{2p}(\xi-z_{0}))-(\phi_{1u}(\infty)-\phi_{1p}(\xi-z_{0}))]^2 \hspace{1cm}     \xi-z_{0}> d_{crit}
\\
\ 0\hspace{10cm} |\xi-z_{0}|< d_{crit} \nonumber
\\ \frac{\textstyle\ MR^2}{\textstyle\ 2}[(\phi_{2u}(-\infty)-\phi_{2p}(\xi-z_{0}))-(\phi_{1u}(-\infty)-\phi_{1p}(\xi-z_{0}))]^2 \hspace{0.5cm} \xi-z_{0}< -d_{crit}
\end{array}   \label{21}
\right.
\end{equation}
\\ The functions $\phi_{ip}$ with $i=1,2$ are the unperturbed functions Eq.(\ref{8}), plus certain induced perturbations by the inhomogeneity. A suitable solutions for $\phi_{ip}$ are possible to find by using the perturbative approach\cite{Eilenberger}. However, due to the extenuant calculus, we propose with good approximation the perturbation given by a function like the shape mode, first, for the relevant contribution that it plays in the interaction dynamics with the inhomogeneity, outlined by some authors in another generic problems\cite{kaberman} and second, because the main effect produced by inhomogeneity in its relative motion towards on solitons in this model is a contraction of soliton radius which is well represented by the shape mode, for being the inhomogeneity a simple bond between the chains, on the contrary of others similar models\cite{Naka1,Naka2,Yo1,Yo2}. Also we consider only the situation of a slow motion for the solitons which does not take into account the deformation due to the velocity.  For the case of antisymmetric functions we get the following potential function:

\begin{equation}
\ V(\xi)=\left\{\begin{array}{c}
2MR^2\left[2\pi-4\arctan\left[\exp\left[\frac{\textstyle\sqrt{2}}{\textstyle\Lambda}(\xi-z_{0})\right]\right]-\frac{\textstyle\beta\sqrt{2}}{\textstyle\Lambda}\frac{\textstyle\sinh(\beta\frac{\textstyle\sqrt{2}}{\textstyle\Lambda}(\xi-z_{0}))}{\textstyle\cosh^{2}(\beta\frac{\textstyle\sqrt{2}}{\textstyle\Lambda}(\xi-z_{0}))}\right]^{2} \hspace{0.1cm}\xi-z_{0}> d_{crit}
\\
\ 0 \hspace{12cm} |\xi-z_{0}|< d_{crit}
\\ 2MR^2\left[4\arctan\left[\exp\left[\frac{\textstyle\sqrt{2}}{\textstyle\Lambda}(\xi-z_{0})\right]\right]+\frac{\textstyle\beta\sqrt{2}}{\textstyle\Lambda}\frac{\textstyle\sinh(\beta\frac{\textstyle\sqrt{2}}{\textstyle\Lambda}(\xi-z_{0}))}{\textstyle\cosh^{2}(\beta\frac{\textstyle\sqrt{2}}{\textstyle\Lambda}(\xi-z_{0}))}\right]^{2}\hspace{0.9cm} \xi-z_{0}< -d_{crit}
\end{array}   \label{22}
\right.
\end{equation}
\\ where $\beta=\frac{\textstyle\ M}{\textstyle\ k}$ is a parameter that we consider gives a measure of the deformation due to the strength of the inhomogeneity.  
\\In order to clarify our results, we will proceed to compare Eq.(\ref{21}) with a potential function calculated by standard methods\cite{Kivshar1,Kivshar2}.
\\ In our model the stronger bonds distributed along the chain correspond to static inhomogeneities. The static inhomogeneities have been considered as perturbations in many nonlinear models\cite{salerno,Kivshar1,Kivshar2}. By using a perturbative procedure we can estimate the effect of inhomogeneities on solitons dynamics if the characteristic size of the impurities is either much larger or much smaller than the soliton size. For take into account only the zeroth-order approximation we suppose that the influence of the inhomogeneities on the soliton gives rise only to a change in their parameters, like the center position, velocity, ect. It is usually named adiabatic approach\cite{Kivshar1}.
\\ For the Eqs(\ref{todas7}), taking the antisymmetric solutions, the coupled equations become:
\begin{equation}
\ \frac {1}{c^{2}}\frac{\partial^{2}\phi}{\partial\ t^{2}}-\frac{\partial ^{2}\phi}{\partial\ z^{2}} +\frac{2}{\Lambda^{2}}\sin\phi=-\frac{2Q}{\Lambda^{2}}\delta\left(z-z_{0}\right)\sin\phi    \label{23}
\end{equation}

\ This is the perturbed Sine-Gordon equation which for $Q=0$ has a
kink-like solution given by:

\ Among the quantities that are conserved during the evolution there is the momentum

\begin{equation}
\ P=-\int_{-\infty}^{\infty} dz \frac{\partial\phi}{\partial\ t}\ \frac{\partial\phi}{\partial\ z}    \label{24}
\end{equation}

\ In the presence of inhomogeneities the momentum is no longer conserved by using Eq(4.6) it is possible to show that it varies according to\cite{Kivshar1,Kivshar2,kaberman,sanchez,scott}:

\begin{equation}
\ \frac{dP}{dt}=-\int_{-\infty}^{\infty} dz \frac{2Q}{\Lambda^2}\delta\left(z-z_{0}\right) \frac{\partial\left(\cos\phi\right)}{\partial\ z}   \label{25}
\end{equation}

\ The adiabatic theory is defined by the assumption that for small enough  values Q, the kink shape will not be affected and its coordinate $xi$ and its velocity $v$ will become slowly changing functions of time. Within this hypothesis it can be shown that, in the non-relativistic limit $v^{2}\ll\ c^{2}$, the kink center obeys the following evolution law.

\begin{equation}
\ \frac{d^{2}\xi}{dt^{2}}=-\int_{-\infty}^{\infty} dz \frac{2Q}{\Lambda^2}\delta\left(z-z_{0}\right) \frac{\partial\left(\cos\phi\right)}{\partial\ z}   \label{26}
\end{equation}

\ where the potential function is

\begin{equation}
\ U\left(\xi\right)=\frac{4 Q}{\Lambda^{2}\cosh^{2}\left(\sqrt{2}\left(\xi-z_{0}\right)/\Lambda\right)} \label{27}
\end{equation}
\ and $\xi$, $z_{0}$ are the positions of the mass center and the inhomogeneity respectively.
\\ The Eq.(\ref{27}) is a dimensionless function, multiplying it by the term $Ka^2$ this becomes
\begin{equation}
\ U\left(\xi\right)=\frac{4 MR^{2}}{\cosh^{2}\left(\sqrt{2}\left(\xi-z_{0}\right)/\Lambda\right)} \label{28}
\end{equation}

\ By comparing both Eq.(\ref{22}) and Eq.(\ref{28}) it is easy to(see Fig.2)that there exist a great differences between the profiles of the two curves.
\\
\ Such differences are explained, on the one hand, because the calculation of the standard potential was done with the zeroth-order approach, which only produces changes in parameters like center position, velocity, without altering the shape. On the other hand, our proposal is a renormalized function which takes with a good approximation the deformation due to the interaction with the inhomogeneity.
However the great difference owes its cause to the fact that the inhomogeneities on the contrary of other models, are localized in the bonds between the chains, therefore when the solitons (rupture of bonds in a localized region along the coupled chains)are situated from the inhomogeneity below the critical distance it is removed because the bond is broken. From the energetic analysis and taking into a account this last issue we derive two implications, first, the maxima elastic potential energy is reached for this critical distance and second,
a potential well is formed whose width is given by twice the critical distance, such as is depicted in the (Fig.\ 2). The existence of this potential well could given place to interesting
dynamics for example localized traps for the solitons distributed along the couple chains. Beside DNA being a low dimensional system these results could give many possibilities in its applications for the develop of new nanoelectronic devices as reported for such systems\cite{Jacob1,Jacob2}. 
\section{Concluding Remarks}

\ We have extended the model of Yakushevich\cite{yaku1} for the DNA torsional dynamics, to take into account the presence of inhomogeneities in the molecular chains. The stability of certain kind of solitons that can propagate through the molecule, simulating the dynamics of open states, are studied. The obtained results are more general than those reported in the literature\cite{Jorge,yaku1} and include them as particular cases. We have found that for distances larger than 1.606 times the soliton's radius, the effect of the inhomogeneities in the structural stability of the open state can be neglected. Furthermore we analyzed the motion of DNA torsional solitons as the motion of point-particles through a renormalized potential function expressed in terms of the critical distance, which could give interesting dynamics.
\\Our study has been done considering the presence of only one inhomogeneity and for the antisymmetric solitons reported in\cite{Jorge} as exact analytical solutions for the unperturbed model. Subsequent studies can be directed to analyze the behavior of solitons between two inhomogeneities in terms of soliton's radius and the distance between inhomogeneities ratio.
The more general case including a well-determined distribution of those bonds along the double helix will be considered in a future work.
\acknowledgments The author L M. Molina wishes to thanks the fruitful comments of the Professors A. Sanchez from the University Carlos III of Madrid and F.G. Mertens from the University of Bayreuth of Germany.
\section*{appendix}
\ The Eq.(12) is a eigenvalue problem which have a quantum
mechanical analogue:
\begin{equation}
\ \frac {d^{2}\phi}{dz^{2}}-\frac{2m\arrowvert\ E\arrowvert\phi}{\hbar^{2}}+\frac {2b^2\phi}
{\cosh^{2}\left(\sqrt{2}(z-z_{m})/\Lambda\right)}+
2Df(\chi)\delta(z-z_{0})=0 \eqnum{A1}  \label{A1}
\end{equation}

\ being:
\begin{equation}
\ f(\chi)=\left[\frac{2}{\cosh^{2}(\sqrt{2}\chi/\Lambda)}-1\right], \hspace{1cm}  \chi=z_{0}-z_{m} \eqnum{A2}  \label{A2}
\end{equation}

\ where the potential function of the Schr\"odinger problem is:

\begin{equation}
\  U(z)=-\frac{\hbar^2}{2m}\left(\frac {2b^2}{\cosh^{2}\left(\sqrt{2}(z-z_{m})/\Lambda\right)}+            
2Df(\chi)\delta(z-z_{0})\right) \eqnum{A3} \label{A3}
\end{equation}
\\ We propose the following solution:

\begin{equation}
\phi(z)=\left\{\begin{array}{c}
 \ A\frac{\textstyle\exp[-Df(\chi)(z-z_{0})]}{\textstyle\cosh(b(z-z_{m}))} \hspace{1cm}\ z>z_{0}\vspace{0.5cm}
\\ \ A\frac{\textstyle\exp[Df(\chi)(z-z_{0})]}{\textstyle\cosh(b(z-z_{m}))} \hspace{1cm}\ z<z_{0}
\end{array}\eqnum{A4} \label{A4}
\right.
\end{equation}
\ with the normalization condition %\\ \centerline{$
$$A^2\left( \int_{-\infty}^{z_{0}}\frac{\textstyle\exp[2Df(\chi)(z-z_{0})]}{\textstyle\cosh^2(b(z-z_{m}))}dz+
\int_{z_{0}}^{\infty}\frac{\textstyle\exp[-2Df(\chi)(z-z_{0})]}{\textstyle\cosh^2(b(z-z_{m}))}dz\right)=1$$%}\\ 
\ whose eigenvalue is
\begin{equation}
\ E=-\frac{\hbar^2}{2m}[b^2+D^2f(\chi)^2] \eqnum{A5}  \label{A5}
\end{equation}

This solution Eq.(A1) in the intervals $(-\infty,z_{0}-\sigma]$,$[z_{0}+\sigma,+\infty)$ where $\sigma$ can take a very small value. If we put the proposed solution in Eq.(\ref{A1}), integrating in the interval and then tending to the limit $\sigma\rightarrow\ 0$ we can demonstrate the uniqueness of the solution with the constraint:
\begin{equation}
\ f(\chi)>0  \eqnum{A6}   \label{A6}
\end{equation}
\ For degenerated cases,i.e., $D=0$, $b\neq\ 0$ and $D\neq\ 0$, $b=0$, the Eq.(\ref{A1}) reduces to:

\begin{equation}
\ \frac {d^{2}\phi}{dz^{2}}-\frac{2m\arrowvert\ E\arrowvert\phi}{\hbar^{2}}+\frac {2b^2\phi}
{\cosh^{2}\left(\sqrt{2}(z-z_{m})/\Lambda\right)}=0\hspace{1cm}\mbox{   with } \  E<0 \eqnum{A7} \label{A7}
\end{equation}
\ and

\begin{equation}
\ \frac {d^{2}\phi}{dz^{2}}-\frac{2m\arrowvert\ E\arrowvert\phi}{\hbar^{2}}+
2Df(\chi)\delta(z-z_{0})=0 \eqnum{A8}  \label{A8}\hspace{1cm}  \mbox{with } \  E<0, D>0
\end{equation}
\ respectively
\\ The above equations were studied in reference\cite{Fluge}.It is easy to show that our proposed solution Eq.(\ref{A4}) with eigenvalues Eq.(\ref{A5})
reduces to those proposed for the corresponding particular cases, i.e. Eq.(\ref{A7}) and Eq.(\ref{A8}).
\\ By comparison of both Eq.(\ref{12}) and Eq.(\ref{A1}) using the above results we obtain the formulae Eq.(\ref{14})- Eq.(\ref{16}).
\\ For the case $f(\chi)<0$, we would obtain a function potential with a similar shape but now with a reverse Delta function away from the
well formed by the hyperbolic function potential.In this case the stationary states formed in the well of the potential function are not affected and corresponds with the eigenfunction for $D=0$ Eq.(\ref{A7}).
\\ From the condition
\begin{equation}
f(\chi)=f(d_{crit})=0  \eqnum{A9}\label{A9}
\end{equation}
derives a distance starting from which either the well potential introduced by Delta function can have influence on the discrete states formed in the well of the function potential or not(see Fig3).The Fig.(3) is the plot of the potential function(Eq.(A3)) with $z_{m}=0$, $b^2=1/2$ and the rest of parameters that not appear in the figure are set to one for simplicity.

\begin{figure}[tbp]
\caption{Section of coupled chains which depicts two couple strands of DNA with a stronger bond joining them.}
\end{figure}

\begin{figure}[tbp]
\caption{Field Potential created by the inhomogeneity in which the soliton moves as a point-particle, a) Standard potential function(dotted-line), b)Renormalized potential function(continuous-line).}
\end{figure}

\begin{figure}[tbp]
\caption{Potential function for a equation like-Schr\"odinger equation derived from the stability analysis.}
\end{figure}
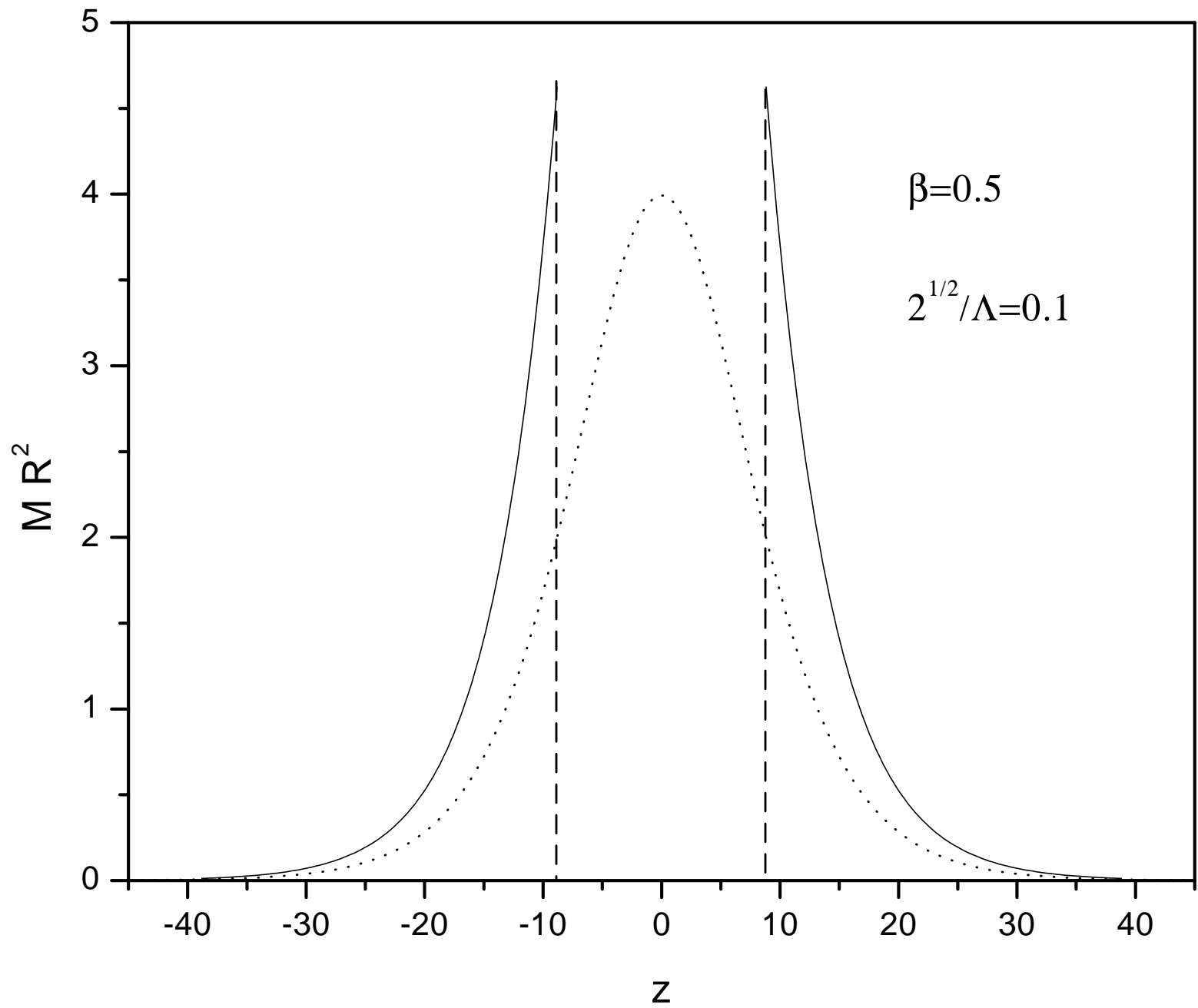
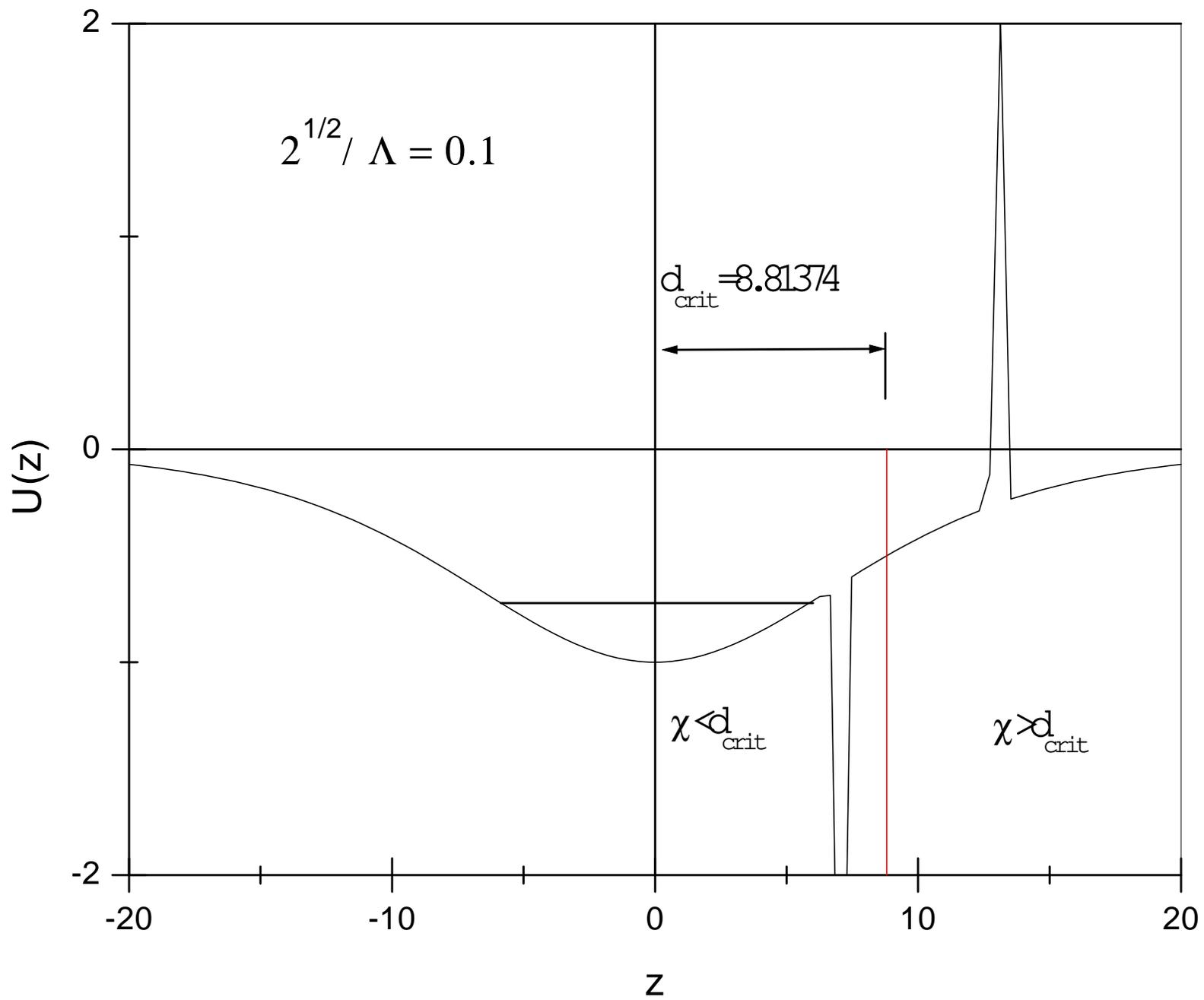
\end{document}